\documentclass[sn-nature]{sn-jnl}% Style for submissions to Nature Portfolio journals
\usepackage{graphicx}%
\usepackage{multirow}%
\usepackage{amsmath,amssymb,amsfonts}%
\usepackage{amsthm}%
\usepackage{mathrsfs}%
\usepackage[title]{appendix}%
\usepackage{xcolor}%
\usepackage{textcomp}%
\usepackage{manyfoot}%
\usepackage{booktabs}%
\usepackage{algorithm}%
\usepackage{algorithmicx}%
\usepackage{algpseudocode}%
\usepackage{listings}%
\usepackage{cleveref}
\usepackage{natbib}
\theoremstyle{thmstyleone}%
\theoremstyle{thmstyletwo}%

\theoremstyle{thmstylethree}%

\begin{document}

\title[Article Title]{CellTypeAgent: Trustworthy cell type annotation with Large Language Models}


\author[1]{\fnm{Jiawen} \sur{Chen}}\email{jiawenn@email.unc.edu}
\equalcont{These authors contributed equally to this work.}

\author[2]{\fnm{Jianghao} \sur{Zhang}}\email{jhzhang@unc.edu}
\equalcont{These authors contributed equally to this work.}

\author*[2,4]{\fnm{Huaxiu} \sur{Yao}}\email{huaxiu@unc.edu}

\author*[1,2,3]{\fnm{Yun} \sur{Li}}\email{yun\_li@med.unc.edu}

\affil[1]{\orgdiv{Department of Biostatistics}, \orgname{University of North Carolina at Chapel Hill}}

\affil[2]{\orgdiv{Department of Computer Science}, \orgname{University of North Carolina at Chapel Hill}}

\affil[3]{\orgdiv{Department of Genetics}, \orgname{University of North Carolina at Chapel Hill}}

\affil[4]{\orgdiv{School of Data Science and Society}, \orgname{University of North Carolina at Chapel Hill}}

% Brief Communication
%A Brief Communication is a concise report describing potentially groundbreaking yet preliminary method or tool developments, highly practical tweaks to an existing method or tool, software platforms, resources of broad interest, and technical critiques of widely used methodologies.

%Format

%Abstract – up to 70 words, unreferenced.
%Main text – 1,200 words (up to 1600 words with editorial discretion), including abstract, references and figure legends.
%The main text should not contain sections nor subheadings. 
%Display items – maximum 2 figures and/or tables (up to 3 items with editorial discretion).
%Online Methods section should be included and should contain subheadings.
%References – as a guideline, we typically recommend up to 20. 
%Brief Communications include received/accepted dates. 
%Brief Communications may be accompanied by supplementary information. 
%Brief Communications are peer reviewed.

%70
\abstract{Cell type annotation is a critical yet laborious step in single-cell RNA sequencing analysis. We present a trustworthy large language model (LLM)-agent, CellTypeAgent, which integrates LLMs with verification from relevant databases. CellTypeAgent achieves higher accuracy than existing methods while mitigating hallucinations. We evaluated CellTypeAgent across nine real datasets involving 303 cell types from 36 tissues. This combined approach holds promise for more efficient and reliable cell type annotation.}


\maketitle

%1,200
\section{Main}\label{sec:intro}

Cell type annotation, serving as an essential first step in single-cell RNA sequencing (scRNA-seq) analysis, demands considerable manual effort from researchers~\citep{hu2023cellmarker,hou2024assessing}. The standard practice involves comparing marker genes expressed in each cell cluster with established marker genes documented in the literature, which is not only time-consuming but also labor-intensive. Advancements in artificial intelligence (AI), particularly in natural language processing (NLP), have introduced Large Language Models (LLMs) like Generative Pre-trained Transformers (GPT)~\citep{achiam2023gpt}. Trained on extensive datasets encompassing a wide range of textual information, LLMs possess the ability to understand and generate human-like language. Their proficiency in pattern recognition and contextual understanding make them powerful tools for various tasks, including those beyond natural language applications~\citep{gong2023evaluating}. Specifically, Hou et al. \cite{hou2024assessing} have assessed the capability of GPT models in automated cell type annotation. Their study demonstrated that GPT not only effectively reduced the manual workload but also outperformed many existing semi-automated and fully automated methods. This highlights the potential of GPT models to revolutionize the cell type annotation process by enhancing both efficiency and accuracy.

Despite the promising capabilities of LLMs, there are prevalent concerns regarding their reliability, particularly the phenomenon known as ``hallucination"~\citep{farquhar2024detecting,huang2025survey}. LLMs are known to occasionally generate responses that are nonsensical or factually incorrect, including fabricating references or providing misleading information. This issue is well-documented and poses significant challenges, especially in critical fields like medicine and biology, where accuracy is paramount.% The propensity for hallucination has made researchers cautious about adopting LLMs in medical domains, as incorrect annotations or interpretations can lead to flawed conclusions and hinder scientific progress.

In this brief communication, we propose a trustworthy LLM agent, CellTypeAgent, specifically designed for cell type annotation using marker genes. We address the challenges associated with LLM hallucinations by integrating GPT models with existing databases. By combining the advanced language understanding capabilities of GPT with real-data verification, we aim to enhance the reliability of cell type annotations while maintaining efficiency.

\begin{figure}[h]
\centering
\includegraphics[width=\textwidth]{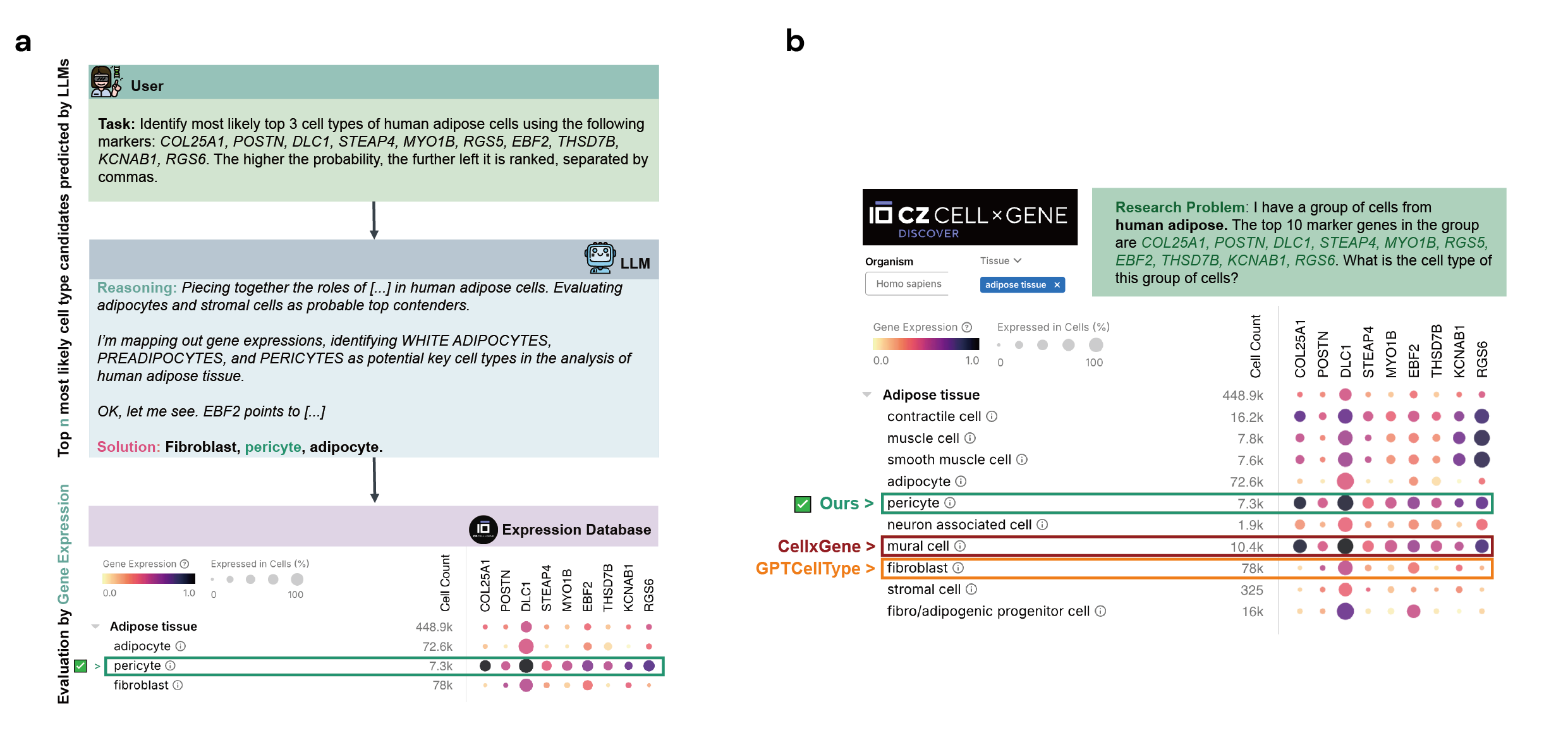}
\caption{CellTypeAgent workflow. (a) CellTypeAgent first suggests several cell type candidates using an LLM. Then the final cell type is decided by cross-referencing with the CellxGene database. (b) A human adipose example of identified final cell type using CellTypeAgent, CellxGene, and GPTCellType. }\label{fig:workflow}
\end{figure}

Our CellTypeAgent conducts an initial inference of potential cell types using an LLM. From the candidates suggested by the model, we then determine the final cell type by cross-referencing with the CellxGene database~\citep{czi2023cz} (\Cref{fig:workflow}a). CellxGene is a comprehensive repository that includes single-cell gene expression data of 1634 datasets from 257 studies, allowing queries based on species (7 species), tissue type (50 tissue types), cell type (over 41 million cells covering 714 cell types), and marker gene name (106,944 genes). By selecting the cell type with the highest average gene expression from this database, we ensure that the final annotation is grounded in empirical data, thereby increasing the trustworthiness of the results.

\begin{figure}[h]
\centering
\includegraphics[width=\textwidth]{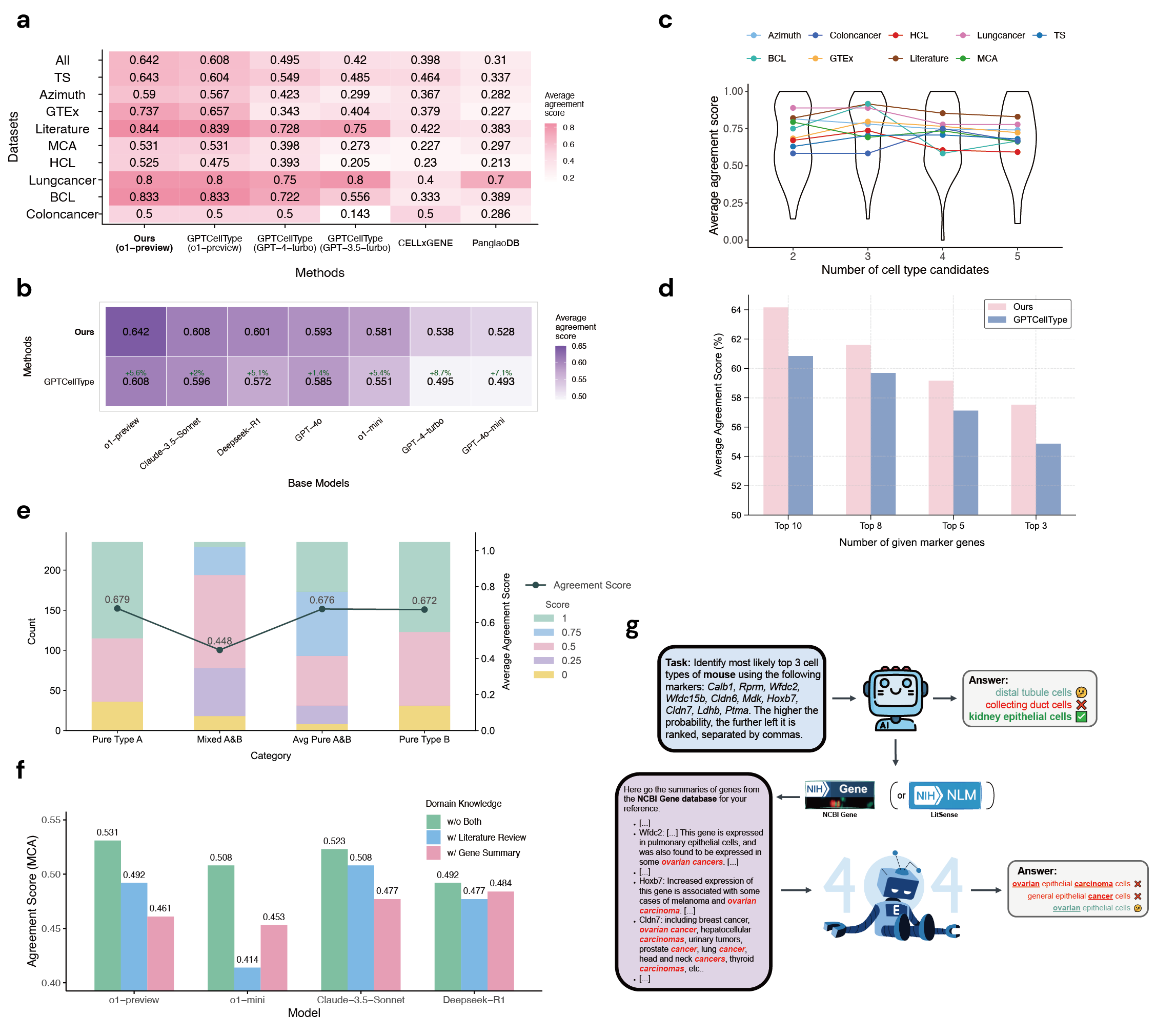}
\caption{CellTypeAgent result. (a) Average agreement score in 9 datasets. We note here that the numerical differences between our GPTCelltype scores and those reported by GPTCelltype are due to updates in the GPT-4 model. All experiments in our study were conducted using GPT models from the May to October 2024 versions. (b) Average agreement score across 9 datasets with various base LLM models. (c) Comparison of performance with different numbers of cell type candidates. (d) Comparison of performance with different numbers of marker genes. (e) Performance with mixed cell type marker genes. Each test is conducted by combining marker genes of two cell types A and B. Pure type A/B is the performance of cell type annotation using marker genes of A/B only. Mixed A\&B is the performance of cell type annotation using marker genes of A and B. Avg pure A\&B is the average performance of pure type A and pure type B in each test. (f) Comparison of performance with and without literature review and gene summary. (g) An example illustrating LLM's heavy reliance on inputting text.}\label{fig:result}
\end{figure}

We systematically evaluated the performance of CellTypeAgent across nine datasets (\Cref{fig:result})~\citep{han2020construction,han2018mapping,the2022tabula,liu2023single,lee2020lineage,kim2020single,eraslan2022single}. For comparison, we assessed its accuracy against competing methods, including GPTCelltype~\citep{hou2024assessing}, CellxGene alone~\citep{czi2023cz}, and another database, PanglaoDB~\citep{franzen2019panglaodb}. Cell type annotation accuracy was evaluated using manual annotations from the original studies as the benchmark (\Cref{sec:Appendix_celltypeagent_method}). Our findings reveal that CellTypeAgent consistently outperforms other methods in all datasets (\Cref{fig:result}a). We further examined CellTypeAgent’s performance using various base LLM models, which provide the initial cell type candidate recommendations (\Cref{fig:result}b). Among these models, the o1-preview model achieved the highest accuracy, suggesting that stronger base models lead to improved annotations. We also evaluated CellTypeAgent against GPTcelltype across different base models and consistently observed superior performance from our method. Notably, when using a weaker base model (treating GPTcelltype as an LLM-only approach), CellTypeAgent demonstrated an even more pronounced improvement, underscoring the benefits of integrating a real database for supporting evidence when the model itself is less robust.

In addition to model capability, data privacy is a common concern when using closed-source LLMs like ChatGPT~\citep{neel2023privacy}. To address this, we tested CellTypeAgent with Deepseek-R1~\citep{guo2025deepseek}, an open-source LLM. Our results (Figure 2b) showed that CellTypeAgent improved the performance by 5.1\%, bringing it very close to the top-performing base models. This gain even surpassed the performance achieved by GPT-4o under GPTcelltype, where GPT-4o outperformed Deepseek-R1 , illustrating that open-source models can deliver competitive results while alleviating privacy concerns with the help of database verification.

We found that the initial inference by the LLM is an essential step in accurate cell type annotation. When comparing CellTypeAgent to using CellxGene alone for inference (i.e., without the initial LLM step), we observed that the CellxGene-only approach exhibited suboptimal performance across most datasets (\Cref{fig:result}a). As an illustrative example, we consider annotating pericyte cells in a human adipose tissue (\Cref{fig:workflow}b). Querying CellxGene yielded multiple cell types (mural cells, pericytes, and muscle cells) exhibiting similarly high average gene expression, leading to ambiguity and frequent misclassification (for example, classifying pericytes as mural cells in this example). In contrast, the LLM component of CellTypeAgent proposes initial candidate cell types based on its learned knowledge from literature. The top 3 candidates in this example are fibroblast, pericyte cell, and adipocyte. This guided inference significantly enhances the accuracy of cell type annotation by effectively narrowing down the most probable cell types for verification. Furthermore, the subsequent verification based on average gene expression correctly identifies pericyte cell, in contrast to GPTCellType’s misclassification as fibroblasts.

We further assessed various factors that could influence CellTypeAgent's performance. First, we examined how the number of initial candidate cell types in the inference step affected accuracy (\Cref{fig:result}c). Although performance remained relatively stable, using the top three candidates yielded a slight higher performance. We also evaluated the impact of the number of marker genes, finding that including more genes generally enhanced annotation quality (\Cref{fig:result}d). Moreover, we tested CellTypeAgent’s ability to handle mixtures of different cell types (\Cref{fig:result}e). When explicitly prompted that multiple cell types might be present, the agent successfully identified one or more components within the mixed sample. While performance declined compared to annotating pure cell types, CellTypeAgent still demonstrated the capacity to accurately detect multiple cell types.

In addition to leveraging real databases like CellxGene, we explored the use of literature search engines (for example, LitSense~\citep{allot2019litsense}) for real data verification (\Cref{fig:result}f,g). Extracting relevant information based on marker genes from literature sources proved to be quite challenging (\Cref{fig:result}f). Our investigation revealed that many papers do not report marker genes for specific cell types within the main text; instead, this information is often located in figures or supplementary tables. Furthermore, when we incorporated literature-extracted information into the initial inference by LLMs, the models tended to rely heavily on the provided text and overlooked their inherent knowledge base (\Cref{fig:result}g), leading to suboptimal performance. A similar issue arose when including gene summaries extracted from the NCBI Gene database, ultimately harming the performance CellTypeAgent  (\Cref{fig:result}f). Additionally, embedding these lengthy literature details within LLM prompts largely increased computational costs, as extended prompts incur higher fees.

In this study, we propose a trustworthy LLM-based cell type annotation tool. By combining LLM-driven inference with real-database verification, our proposed CellTypeAgent significantly enhances cell type annotation in scRNA-seq data. The LLM’s initial inference effectively narrows down possible cell types, while the subsequent verification step reduces errors stemming from model hallucinations. CellTypeAgent remains adaptable and consistently outperforms both simple database-only methods and LLM-only approaches. Future work may further refine how external literature sources are integrated, while also exploring broader applications of this hybrid strategy to ensure both accuracy and scalability.

\bmhead{Acknowledgements}

J.C is supported by National Institutes of Health R01AG079291, RF1-AG082938, R01AG085581, R01AR083790. CellTypeAgent and all code that could be used to reproduce the analysis are publicly available at \url{https://github.com/jianghao-zhang/CellTypeAgent/}.

\begin{appendices}

\section{CellTypeAgent method}\label{sec:Appendix_celltypeagent_method}

We present \textbf{CellTypeAgent}, a reliable and efficient framework that automatically annotates and evaluates cell types using LLMs powered by \textit{biomedical knowledge tools}. Our approach encompasses two complementary stages:

\subsection{Stage 1: LLM-based Candidates Prediction}

Let $\mathcal{G} = \{g_i\}_{i=1}^n$ denote a set of marker genes from tissue $\tau \in \mathcal{T}$ of species $\mathcal{s} \in \mathcal{S}$. We employ advanced LLMs to generate an ordered set of cell type candidates. The prompt we used is formatted as ``Identify most likely top 3 celltypes of (tissue type) using the following markers: (marker genes). The higher the probability, the further left it is ranked, separated by commas.". See \Cref{fig:workflow} for an example.

%$\mathcal{C} = \{c_i\}_{i=1}^n, \quad c_i \in \Omega$, where $\Omega$ represents the universe of possible cell types. We adjust the prompt to constrain the LLMs to rank candidates by their conceptual probability, formalized as: $P(c_i|\mathcal{X}) > P(c_j|\mathcal{X}), \quad \forall i < j$

\subsection{Stage 2: Gene Expression-Based Candidate Evaluation}

We leverage extensive quantitative gene expression data from CZ CELLxGENE Discover~\cite{czi2023cz} to evaluate the candidates and select the most convinced one. Given species $\mathcal{s}$ and/or tissue $\tau$, we extract the expression data including expression value $e(\mathcal{s}, \tau)$ and expressed ratio $\rho(\mathcal{s}, \tau)$. Then for each gene $g \in \mathcal{G}$ in cell type $c \in \mathcal{C}$, we denote:

\begin{itemize}
    \item $e_{gc}(\mathcal{s},\tau)$: scaled expression value of gene $g$ in cell type $c$
    \item $\rho_{gc}(\mathcal{s}, \tau)$: expressed ratio of gene $g$ in cell type $c$
\end{itemize}

In stage 1, we utilize LLM for initial candidate ranking. In order to use the ranking information in this stage, we need to assign initial rank scores to the candidates. The initial rank score $r_i$ for candidate $c_i$ is defined as: $r_i = n - i, \quad i \in \{1,\ldots,n\}$. For any value set $\mathcal{V}$, the $\text{rank}$ function is defined as: $\text{rank}(v) = |\text{unique}(\mathcal{V})| - i - 1, \quad \text{where } v \text{ belongs to the } i\text{-th} \text{ distinct value in } \mathcal{V}$.

We introduce the selection score function $\text{score}: \mathcal{C} \rightarrow \mathbb{R}$ as follows:

\begin{itemize}
\item When tissue type $\tau$ is known:

$\text{score}(c) = r_c + \text{rank}\left(\sum_{g}e_{gc}(\mathcal{s},\tau)\right) + \text{rank}\left(\sum_{g}\rho_{gc}(\mathcal{s},\tau)\right) + \frac{1}{|\mathcal{T}|}\sum_{\tau}\text{rank}\left(e_{gc}(\mathcal{s})\right)$.

\item When tissue type $\tau$ is unknown:

$\text{score}(c) = r_c + \text{rank}\left(\sum_{\tau}\sum_{g}e_{gc}(\mathcal{s})\right) + \text{rank}\left(\sum_{\tau}\sum_{g}\rho_{gc}(\mathcal{s})\right) + \frac{1}{|\mathcal{T}|}\sum_{\tau}\text{rank}\left(\sum_{g}e_{gc}(\mathcal{s})\right)$.
\end{itemize}

Then the optimal cell type annotation $c^*$ within $\mathcal{C}$ is determined by:

$$c^* = \operatorname{argmax}_{c \in \mathcal{C}} \text{score}(c)$$

The first term $r_c$ incorporates the LLM's logical reasoning based on both its internal knowledge and/or information from external literature and gene databases. The second and third terms reflect the gene's specific quantitative expression patterns within the given cell type, tissue, and species group. The last term summarizes the gene's overall quantitative expression profile for that cell type across all tissues of that species.

\subsection{Optional literature search and gene summary}

For each gene $g_i$, we use two optional tools to retrieve complementary information:

\begin{itemize}
    \item $\mathcal{L}(g_i)$: relevant literature segments retrieved from LitSense
    \item $\mathcal{I}(g_i)$: gene information from NCBI gene database
\end{itemize}

The composite input for LLM-based prediction is formalized as: $\mathcal{X} = \{\mathcal{s}, \tau, \mathcal{G}\} \cup \mathcal{O}$, where $\mathcal{O} \subseteq \{\{\mathcal{L}(g_i)\}_{i=1}^n, \{\mathcal{I}(g_i)\}_{i=1}^n\}$ is the optional knowledge.

\section{Evaluation metrics}

We evaluate our method using the agreement score metric proposed in GPTCellType~\citep{hou2024assessing}. For each predicted cell type, we first map both the prediction and ground truth annotation to standardized Cell Ontology (CL) terms and broad cell type categories. The agreement between predicted and manual annotations is then classified into three levels:

\begin{itemize}
    \item \textbf{Full Match} (score = 1): Identical annotation terms or CL ontology names
    \item \textbf{Partial Match} (score = 0.5): Same or hierarchically related broad cell type names (e.g., fibroblast vs. stromal cell) but different specific annotations
    \item \textbf{Mismatch} (score = 0): Different broad cell types, specific annotations, and CL terms
\end{itemize}

The final agreement score for each dataset is computed as the average across all samples:

$$\text{Agreement Score} = \frac{1}{|\mathcal{D}|}\sum_{d \in \mathcal{D}} s_d,$$

where $\mathcal{D}$ is the set of all predictions in the dataset and $s_d \in \{0, 0.5, 1\}$ represents the agreement score for each prediction.

\section{Mixed cell type evaluation}

To assess the performance of CellTypeAgent when the marker genes are from mixed cell types, we conducted the following experiment. We manually created pseudo-mixed cell types by combining the marker genes of two distinct cell types, resulting in 235 test cases. We then added the following instruction to the prompt used in our standard \textit{CellTypeAgent} pipeline: ``Note that the sample could be a mixture of multiple cell types. If you detect markers from different cell types, please indicate it as a mixture in your prediction.''

Here is an illustrative example of the final top-3 candidate predictions (after processing):

\verb|[['endothelial cell', 'T cell'], ['endothelial cell'], ['T cell']]|.

In this format, each sub-list represents one candidate. If a sub-list contains multiple cell types (e.g., \verb|['endothelial cell', 'T cell']|), this indicates a mixed prediction.

We restructured our evaluation for mixed prediction. Originally, if a candidate was a single cell type (e.g., \verb|['endothelial cell']|), we simply sorted the candidates based on their individual scores. In the case of a mixed candidate (e.g., \verb|['endothelial cell', 'T cell']|), we now compute the average of the two cell types’ scores and then compare this average to other candidates during ranking.

In contrast to our prior evaluation method, each sample’s ground truth can now be a list of cell types (e.g., \verb|['A', 'B']|). Consider the following set of candidate predictions for a given sample: \verb|[['A', 'b'],['B'],['D']]|.

\begin{itemize}
    \item For the first candidate, \verb|['A', 'b']|, each predicted cell type is compared with each ground-truth cell type to derive an agreement score. We then choose the permutation of matches that maximizes the cumulative score and average those matches to obtain the final score. For instance, matching \verb|'A'| to \verb|'A'| yields 1, and \verb|'b'| to \verb|'B'| yields 0.5, producing a final score of $(1 + 0.5)/2 = 0.75$.
    \item For the second candidate, \verb|['B']|, we compare \verb|'B'| against both \verb|'A'| and \verb|'B'|, then average those scores. In this example, that would be $(0 + 1)/2 = 0.5$.
    \item For the third candidate, \verb|['D']|, comparing \verb|'D'| with both \verb|'A'| and \verb|'B'| yields scores of 0 and 0, respectively, resulting in a final score of 0.
\end{itemize}

\section{Evaluation dataset}

We collected the cell type marker genes and the annotated cell types from literature~\citep{han2020construction,han2018mapping,the2022tabula,liu2023single,lee2020lineage,kim2020single,eraslan2022single}. For the HuBMAP Azimuth project, we downloaded the manually annotated cell types and their marker genes from the Azimuth website (\url{https://azimuth.hubmapconsortium.org/}). We utilized the datasets evaluated in \cite{hou2024assessing}. See \cite{hou2024assessing} for a more detailed summary.

\end{appendices}


%\bibliography{ref}% common bib file


\end{document}